 \documentclass[final,5p,times,twocolumn]{elsarticle}
 \usepackage{graphicx}
\usepackage{bm}
\usepackage{txfonts}
\usepackage{hyperref}
\usepackage{mathrsfs}
\usepackage[english]{babel}
\usepackage{amsmath}
\usepackage{cleveref}
\usepackage[autostyle, english = american]{csquotes}
\MakeOuterQuote{"}

\date{\today}
\newcommand{\be}{\begin{eqnarray}}
	\newcommand{\ee}{\end{eqnarray}}

\newcommand{\bfz}{{\bf 0}_{\perp}}

\newcommand{\bfk}{{\bf k}_{\perp}}

\newcommand{\bfkj}{{\bf k}_{\perp j}}

\newcommand{\bfP}{{\bf P}_{\perp}}

\usepackage{xcolor}

\usepackage{amssymb}
\usepackage{lipsum}

\journal{Physics Letters B}

\begin{document}

\begin{frontmatter}




\title{Does nuclear medium affect the transverse momentum-dependent parton distributions of valence quark of pions ?}

 \author{Navpreet Kaur}
	\ead{knavpreet.hep@gmail.com}

\author{Satyajit Puhan}
\ead{puhansatyajit@gmail.com}
\author{Reetanshu Pandey}
\ead{reetanshuhep@gmail.com}
\author{Arvind Kumar}
\ead{kumara@nitj.ac.in}
\author{Suneel Dutt}
\ead{dutts@nitj.ac.in}
\author{Harleen Dahiya}
\ead{dahiyah@nitj.ac.in}
	\affiliation{addressline={Computational High Energy Physics Lab, Department of Physics}, 
		city={Jalandhar}, 
		organization={Dr. B.R. Ambedkar National
			Institute of Technology}, state={Punjab},postcode={144008}, 
		           country={INDIA}}

\begin{abstract}
We calculate the valence quark transverse momentum-dependent parton distributions (TMDs) of the lightest pseudoscalar meson, pions, in isospin asymmetric nuclear matter at zero temperature by employing a light-cone quark model.
The medium modifications in the pion unpolarized TMDs are induced through the effective  quark masses computed using the chiral SU($3$) quark mean field model. The spin densities at different momentum fraction ($x$) have also been calculated at different baryonic densities.

\end{abstract}




\end{frontmatter}




\section{Introduction} 
\label{introduction}
In recent years, the study of partonic structure of hadrons has been an interesting topic for both experimental and theoretical realms of hadron physics. The understanding of internal structure of hadrons in term of valence quark, gluon and sea quark distributions is still a mystery along with the constituent elements behavior within different nuclear media. The three dimensional transverse momentum-dependent parton
distribution functions (TMDs) \cite{Diehl:2015uka,Angeles-Martinez:2015sea,Puhan:2023ekt} and one dimensional parton distribution functions
(PDFs) \cite{Martin:1998sq} describe the longitudinal and transverse quark distributions inside a hadron. PDFs are the simplest distribution functions having information about longitudinal momentum fraction $(x)$ carried by the constituent quark from parent hadron but lack in the information about the transverse momentum $(\bfk)$ of the quark. PDFs are extracted from the deep inelastic scattering (DIS) processes \cite{Polchinski:2002jw}. The extended form of PDFs with the inclusion of transverse degree of freedom give rise to multi-dimension TMDs. The TMDs carry information about longitudinal momentum fraction along with transverse momentum of the quark. The TMDs encode information about the correlations among spin of the hadron, spin of the parton, and the parton's transverse momentum. Orbital motion, intrinsic transverse motion, spin polarization of the parton inside a hadron are also encoded by TMDs. These can be directly related to measurable quantities in experiments, such as azimuthal asymmetries in semi-inclusive deep inelastic scattering (SIDIS) and Drell-Yan (DY) processes. TMDs are extracted from the DY processes \cite{Zhou:2009jm}, SIDIS processes \cite{Bacchetta:2017gcc} and $Z^0/ W^\pm $ production \cite{Catani:2015vma}.

\par In order to have a glimpse of partonic dynamics, we have chosen to explore an unpolarized T-even TMD that provides an insight into the intrinsic transverse momentum of partons inside a hadron \cite{Meissner:2008ay}. The unpolarized $f_1(x, \bfk^2)$ TMD of pion has been calculated in different non-perturbative models like Nambu-Jona-Lasinio (NJL) model \cite{Noguera:2015iia}, MIT Bag model \cite{Lu:2012hh} and light-front (LF) dynamics \cite{Ahmady:2019yvo,Kaur:2020vkq}, along with theoretical extractions from DY data \cite{Cerutti:2022lmb}. However, all of the above works have been reported without introducing the medium effects. As the European Muon Collaboration (EMC) has already indicated that finite density of nuclear medium may have considerable impact on the internal structure of hadrons \cite{EuropeanMuon:1983wih},
therefore, the analysis of medium modification effects in hadronic and nuclear physics becomes crucial with a comparative interpretation of vacuum and in-medium observables.
It will enhance our understanding to visualize the internal structure of pions in terms of the momentum distribution among quarks inside a hadron, when immersed in a nuclear matter \cite{Brown:1991kk,Lee:2020ywf}. It therefore becomes interesting to study the effect of nuclear medium on the unpolarized T-even $f_1(x, \bfk^2)$ TMD of pion.

\par Very limited work has been reported in the field of nuclear medium modification effects on the valence quark distributions of pion \cite{Suzuki:1995vr,deMelo:2014gea,Hutauruk:2019ipp,Hutauruk:2021kej}. In Ref. \cite{Suzuki:1995vr}, the structure function of pion has been studied in nuclear medium by employing NJL model. Whereas in Ref. \cite{deMelo:2014gea}, electromagnetic form factors, charge radii, and weak decay constants were explored using a combined approach of light-front pion wave function based on a Bethe-Salpeter amplitude model and quark meson coupling (QMC) model. The pion and kaon valence quark (gluon) distributions have also been studied using NJL and QMC model in Refs. \cite{Hutauruk:2019ipp,Hutauruk:2021kej}. 
Along with these works, a variety of models and approaches such as the Linear-sigma
model (L$\sigma$M) \cite{Suenaga:2019urn}, the instanton
liquid model \cite{Nam:2008xx}, the hybrid light front-quark-meson coupling (LF-QMC) models \cite{deMelo:2016uwj, deMelo:2018hfw},  the Dyson-Schwinger equation
(DSE) based approach \cite{Fritzsch:1999ee} and the QCD sum rules (QSR) \cite{Park:2016xrw, Bozkir:2022lyk} have been employed to study the in-medium properties of light and heavy hadrons.
However, no work has been reported on medium modification of TMDs yet. In our previous work \cite{Puhan:2024xdq}, we found that the asymmetric nuclear matter  can have an  impact of the order of  $10\%$ on the PDFs. Taking motivation from this work, we have extended our work to analyze the effect of baryonic density on transverse momenta of the $u$-quark inside a pion, when immersed in an asymmetric nuclear matter at zero temperature.

\par In this present work, we have used the light-cone quark model (LCQM) \cite{Brodsky:1997de,Qian:2008px,Acharyya:2024enp} for the calculations of vacuum properties, whereas the in-medium effects have been carried out by using the chiral quark mean field (CQMF) model \cite{Wang:2001jw} as well as LCQM. The non-perturbative vacuum unpolarized $u$-quark $f_1(x,\bfk^2)$ pion TMD has been calculated by solving the quark-quark correlator in LCQM \cite{Meissner:2008ay}.
The light-cone concept, upon which LCQM is built, offers an useful framework for visualizing the relativistic implications of partonic dynamics within the hadron \cite{Kaur:2020vkq,Puhan:2023hio}. It is a gauge-invariant, relativistic framework that describes the structure and characteristics of hadrons in a non-perturbative manner. Since they are the main components in charge of the general structure and characteristics of hadrons, valence quarks are the main focus of LCQM.
The physical characteristics of pions, including their electromagnetic properties, charge radii, mass eigen values, decay constants, distribution amplitudes, and PDFs, have been successfully portrayed by LCQM. We have taken the minimal Fock-state of pion i,e, $|\mathcal{P} \rangle =\sum |q \bar{q} \rangle \psi_{q\bar{q}}$ to solve the pion TMD for this work. 
The CQMF model treats the quarks, confined inside baryons through a confining potential, as degrees of freedom. These confined constituent quarks interact through the exchange of scalar field $\sigma, \zeta$ and $\delta$ and the vector fields $\omega$ and  $\rho$ as discussed in Sec. \ref{SecModel}. 

Following is outline of this work: In Sec.
\ref{SecModel} we have presented the details of CQMF and light cone quark model.
The TMD calculations have been detailed in Sec. \ref{Sec_TMDs} with the results in Sec. \ref{Sec_results}. The summary and conclusions are presented in Sec. \ref{Sec_summary}.

\section{The models}
\label{SecModel}

\subsection{Chiral SU(3) quark mean field model}
\label{Sec_CQMF}
In the present section, we have presented  the essential details of the CQMF model through which the medium modifications in the pion TMDs are induced in LCQM. This model incorporates the chiral symmetry and its spontaneous breaking: the low energy characteristics of QCD. These properties arise by considering the Lagrangian densities involving scalar-isoscalar fields $\sigma$ and $\zeta$, along with a scalar-isovector field $\delta$. The scalar-isoscalar fields are responsible for the interactions among quarks in the bound state of a baryon, whereas the scalar isovector field comes into action when there is a finite isospin asymmetry in the medium. The dilaton field $\chi$ is also considered in order to involve the broken scale invariance property of QCD. The exchange of scalar ($\sigma$, $\zeta$ and $\delta$ ) and vector ($\omega$ and $\rho$) fields induce alteration in the  properties of the baryon and its constituent quark flavors $q$. More specifically, effective masses $m_q^\ast$ and energies $e_q^\ast$ are the consequences of the presence of scalar and vector potentials in the medium respectively. At finite temperature $T$, the thermodynamic potential for the isospin asymmetric nuclear matter is given as \cite{Wang:2001jw}
%

\begin{eqnarray}
	\Omega &=& - \sum_{i} \frac{\gamma_i}
	{48\pi^2}  \Big[ \left(2 k_{Fi}^3 - 3M_i^{*2}k_{Fi}\right)\nu_i^{*} 
	\nonumber \\
	&+&
	3 m_i^{*4} \text{ln} \left(\frac{k_{Fi} + \nu_{i}^*}{m_i^*}\right) \Big] 
	 -{\cal L}_{M}-{\cal V}_{\text{vac}}.
	\label{Eq_therm_pot1}  
\end{eqnarray}
Here, the summation is over the nucleons of the medium, i.e., $i = p, n$ and the degeneracy factor $\gamma_i$ takes the value $2$. Also, $k_{Fi}$ is the Fermi momentum of the nucleon. The effective chemical potential $\nu_i^\ast$ of the nucleons is defined in terms of free chemical potential $\nu_i$ as $\nu_i^\ast = \nu_i - g_{\omega}^i\omega -g_{\rho}^i I^{3i} \rho.$  Zero vacuum energy can also be achieved by subtracting the vacuum potential energy, ${\cal V}_{\text{vac}}$. An effective energy $E^{\ast }(k)$ can be expressed in terms of its effective mass as $E^{\ast }(k)=\sqrt{M_i^{\ast 2}+k^{2}}$ and the effective mass $M_i^\ast$ of a baryon in terms of its spurious center of momentum $\langle p_{i~cm}^{\ast 2} \rangle$ \cite{Barik:2013lna} and $e_q^\ast$ can be related through a relation 
\begin{eqnarray}
	M_i^\ast = 
	\sqrt{\biggl(\sum_q n_{qi} e_q^\ast + E_{i~spin} \biggr)^2  - \langle p_{i~cm}^{\ast 2} \rangle} \, ,
\end{eqnarray}
where number of $q$ flavored quark(s) in the $i^{th}$ kind of baryon is denoted by $n_{qi}$. The term $E_{i~spin}$, which plays the role of correction term to baryon energy due to spin-spin interaction, is adjusted to fit the baryon vacuum mass. \par
The term ${\cal L}_{M}$ expressed in Eq. (\ref{Eq_therm_pot1}) involves the contribution of 
${\cal L}_{X}$, ${\cal L}_{VV}$ and ${\cal L}_{\chi SB}$. The term ${\cal L}_{X}$ corresponds to the self-interactions of scalar mesons and the dilaton field and can be expressed as 
\begin{eqnarray}
	{\cal L}_{X} &=& -\frac{1}{2} \, k_0\chi^2
	\left(\sigma^2+\zeta^2+\delta^2\right)+k_1 \left(\sigma^2+\zeta^2+\delta^2\right)^2
	\nonumber \\ 
	&+&k_2\left(\frac{\sigma^4}{2} +\frac{\delta^4}{2}+3\sigma^2\delta^2+\zeta^4\right)+ k_3\chi\left(\sigma^2-\delta^2\right)\zeta \nonumber \\ 
	&-& k_4\chi^4-\frac14\chi^4 {\rm ln}\frac{\chi^4}{\chi_0^4} +
	\frac{\xi}
	3\chi^4 {\rm ln}\left(\left(\frac{\left(\sigma^2-\delta^2\right)\zeta}{\sigma_0^2\zeta_0}\right)\left(\frac{\chi^3}{\chi_0^3}\right)\right). \nonumber \\ \label{scalar0}
\end{eqnarray}
The last two logarithmic terms in the above equation address the scale breaking effects and facilitate in expressing the trace of energy momentum tensor proportional to the fourth power
of the dilaton field $\chi$ within the CQMF model.
On the other hand, the terms ${\cal L}_{VV}$ and ${\cal L}_{\chi SB}$ correspond to the self-interactions of vector mesons and the explicit symmetry breaking respectively. These Lagrangian densities can be expanded in terms of their contributing fields as
\begin{eqnarray}
	{\cal L}_{VV} =&\frac{1}{2} \, \frac{\chi^2}{\chi_0^2} \left(m_\omega^2\omega^2+m_\rho^2\rho^2\right)  + g_4\left(\omega^4+6\omega^2\rho^2+\rho^4\right) \, , 
	\label{vector}
\end{eqnarray}
and 
\begin{equation}\label{L_SB}
	{\cal L}_{\chi SB}=\frac{\chi^2}{\chi_0^2}\left[m_\pi^2\kappa_\pi\sigma +
	\left(
	\sqrt{2} \, m_K^2\kappa_K-\frac{m_\pi^2}{\sqrt{2}} \kappa_\pi\right)\zeta\right] \, .
\end{equation}
Analogous to the scalar-isovector field $\delta$ in ${\cal L}_{X}$, the vector-isovector field $\rho$ is also involved in ${\cal L}_{VV}$ when there is finite isospin asymmetry in the medium. At one loop level, for three colors and three flavors, QCD $\beta$-function defines the order of parameter $\xi$ \cite{Papazoglou:1998vr}. In CQMF model, as a baryon is supposed to be a bound state of quarks with a confining potential, $ \chi_{c}(r)=\frac14 k_{c} \, r^2(1+\gamma^0)$ and interact through the exchange of fields. Thus, one can define the Dirac equation for the quark field $\Psi_{qi}$ as
\begin{equation}
	\left[-i\vec{\alpha}\cdot\vec{\nabla}+\chi_c(r)+\beta m_q^*\right]
	\Psi_{qi}=e_q^*\Psi_{qi}. \label{Dirac}
\end{equation}
In terms of scalar fields ($\sigma, \zeta$ and $\delta$), the effective quark mass $m_{q}^\ast$ is expressed through the relation
\begin{equation}
	m_q^*=-g_\sigma^q\sigma - g_\zeta^q\zeta - g_\delta^q I^{3q} \delta + \Delta m \, . \label{qmass}
\end{equation}
Here, $\Delta m$ is taken to be $77$ MeV for strange $s$ quark. However, for non-strange $u$ and $d$ quarks, it becomes zero. The values of a parameter $k_c$, defined in the expression of confining potential and couplings are fitted to the binding energy $-16$ MeV at nuclear saturaton density $0.16$ fm$^{-3}$ \cite{Kumar:2023owb}. In order to get a reasonable value for the vacuum mass of $s$ quark ($m_s = 450$ MeV), the term $\Delta m$ is incorporated through the Lagrangian density   ${\cal L}_{\Delta m} = - (\Delta m) \bar \psi S_1 \psi$,  where  $S_1 \, = \, \frac{1}{3} \, \left(I - \lambda_8\sqrt{3}\right) = {\rm diag}(0,0,1)$ is the matrix for strange quark \cite{Singh:2016hiw}.
Effective energy of a specific quark, subjected to the meson field is written as
\begin{equation} 
	e_q^*=e_q-g_\omega^q\omega-g_\rho^q I^{3q}\rho\,.
	\label{eq_eff_energy1}
\end{equation}
The thermodynamic potential defined in Eq. (\ref{Eq_therm_pot1}) is minimized with respect to the scalar fields
($\sigma$, $\zeta$, $\delta$,
and $\chi$) and the vector fields ($\omega$ and $\rho$) 
through
\begin{eqnarray}
	\frac{\partial \Omega}{\partial \sigma} = 
	\frac{\partial \Omega}{\partial \zeta} =
	\frac{\partial \Omega}{\partial \delta} =
	\frac{\partial \Omega}{\partial \chi} =
	\frac{\partial \Omega}{\partial \omega} =
	\frac{\partial \Omega}{\partial \rho}  =
	0 \, .
	\label{eq:therm_min1}
\end{eqnarray}
For finite baryon density and isospin asymmetry of the matter, the system of non-linear equations derived for the scalar and vector fields are solved. The isospin asymmetry of the nuclear matter is defined in terms of isospin asymmetry parameter,  $\eta = \frac{\rho_n - \rho_p}{2\rho_{B}}$ with $\rho_B = \rho_p + \rho_n$ as the total baryonic density. 


\subsection{Light-cone quark model}
The expansion of baryon eigenstate $|\mathcal{M}(P^+,\bfP,S_z)\rangle$
for a hadron having total momentum $P$ with light-cone coordinates $(P^+,P^-,\bfP)$ and longitudinal spin projection $S_z$ can be expressed in terms of multiparticle Fock eigenstates $|n\rangle$ as \cite{Qian:2008px,Lepage:1980fj}
\be
|\mathcal{M}(P^+,\bfP,S_z)\rangle= \sum_{n,\lambda_j} \int \prod_{j=1}^{n} \frac{dx_j~  d^2\bfkj}{2(2\pi)^3\sqrt{x_{j}}} \, 16 \pi^{3} \, \nonumber \\
 \delta ~ \bigg(1-\sum_{j=1}^{n} x_{j}\bigg) \, \delta^{(2)} \bigg(\sum_{j=1}^{n}\bfkj\bigg) \nonumber \\		
 \psi_{n/\mathcal{M}}(x_{j},\bfkj,\lambda_{j})|n; x_{j} P^{+},x_{j}\bfP + \bfkj,\lambda_{j}\rangle \, ,
\label{MesonState}\ee
where $x_j=\frac{k_j^+}{P^+}$ with $0 \leq x_j \leq 1$ is the longitudinal momentum fraction of the $j$th constituent parton with $\bfkj$ and $\lambda_j$ as its  transverse momentum and helicity respectively. 
The multiparticle state of $n$-particles is normalized as 
\be
\langle n; k^{\prime +}_j, \bfkj^\prime, \lambda_{j}^\prime|n ; k^+_j, \bfkj, \lambda_j \rangle = \prod_{j=1}^{n} 16 \pi^{3} \,  k^{\prime +}_j \, \delta (k^{\prime +}_j-k^+_j) \, \nonumber \\
 \delta^{(2)} ( \bfkj^\prime-\bfkj) \, \delta_{\lambda_{j}^\prime \lambda_{j}} \, .
\ee
The momenta of meson and its constituent quarks $u$($\bar{d}$), having effective  masses $M^*$ and $m_u^*$($m_{\bar{d}}^*$) respectively, in the light-cone frame are expressed as
\be 
P &=& \bigg(P^+,\frac{M^{\ast2}}{P^+},\bfz\bigg) \, ,  \nonumber \\
k_1 &=& \bigg(x_1 P^+,\frac{\bfk^2 + m^{\ast2}_u}{x_1 P^+},\bfk \bigg) \, ,  \nonumber \\
k_2 &=& \bigg(x_2 P^+,\frac{\bfk^2 + m^{\ast2}_{\bar{d}}}{x_2 P^+},-\bfk \bigg) \, . 
\ee
Under the condition $\sum_{j=1}^{n} x_j = 1$, for the light-cone quark momentum of two constituent quarks of a meson, we get $x_1 + x_2 =1$ which implies that if quark carries $x$ fraction of longitudinal momentum, then its anti-quark partner will carry remaining $(1-x)$ longitudinal momentum fraction.
For a pseudoscalar $\pi$ meson with spin $S_z=0$, the two-particle Fock state can be expanded in the form of light-cone wave functions (LCWFs) as
\be 
|\mathcal{\pi} (P^+,\bfP,S_z)\rangle &=& \int \frac{dx \, d^2 \bfk}{  16 \pi^3 \sqrt{x(1-x)}} \, \nonumber \\ &\times& \big[ \psi (x,\bfk,\uparrow,\uparrow) \, |x P^+, \bfk, \uparrow, \uparrow \rangle   \nonumber \\
&+& \psi (x,\bfk,\uparrow,\downarrow) \, |x P^+, \bfk, \uparrow, \downarrow \rangle  \nonumber \\
&+& \psi (x,\bfk,\downarrow,\uparrow) \, |x P^+, \bfk,  \downarrow,\uparrow \rangle \nonumber \\ &+& \psi (x,\bfk,\downarrow,\downarrow) \, |x P^+, \bfk, \downarrow, \downarrow \rangle \big] \, \label{eqnq} .
\ee 
These LCWFs are composed of momentum space $\varphi$ and spin $\Phi$ wave functions as \cite{Huang:1994dy} 
\be 
\psi(x,\bfk,\lambda_1, \lambda_2)= \varphi(x,\bfk) \, \Phi (x,\bfk,\lambda_1, \lambda_2) \, ,
\ee 
where $\lambda_1(\lambda_2)$ is the helicity of the quark (anti-quark).
The spin wave function $\Phi(x,\bfk,\lambda_1, \lambda_2)$ can be obtained by transforming instant-form SU($6$) wave function into light-cone form by making the use of Melosh-Wigner rotation. In the light-cone formalism, they are expressed as
\cite{Xiao:2002iv,Xiao:2003wf}
\begin{eqnarray}
\Phi(x,\bfk,\uparrow,\downarrow)&=&\frac{[(M^\ast x +m^\ast_u) (M^\ast (1-x)+m^\ast_{\bar{d}})-\bfk^2]}{\sqrt{2}c_1 c_2}  \, , \nonumber \\
\Phi(x,\bfk,\downarrow,\uparrow)&=&\frac{-[(M^\ast x +m^\ast_u) (M^\ast (1-x) +m^\ast_{\bar{d}})-\bfk^2]}{\sqrt{2}c_1 c_2}  \, , \nonumber \\
\Phi(x,\bfk,\uparrow,\uparrow)&=&\frac{[(M^\ast x +m^\ast_u) k_2^l - (M^{\ast} (1-x) +m^\ast_{\bar{d}}) k_1^l]}{\sqrt{2}c_1 c_2}  \, , \nonumber \\
\Phi(x,\bfk,\downarrow,\downarrow)&=&\frac{[(M^\ast x +m^\ast_u) k_2^r - (M^\ast (1-x) +m^\ast_{\bar{d}}) k_1^r]}{\sqrt{2}c_1 c_2}  \, . \nonumber \\
\label{SpinWfns}
\end{eqnarray}  
In the above equations, we have $c_1 = \sqrt{(M^\ast x +m_u^{\ast2})^2 + \bfk^2}$, $c_2 = \sqrt{(M^\ast (1-x) +m_{\bar{d}}^{\ast2})^2 + \bfk^2}$ and $k_{1(2)}^{r,l} = k_{1(2)}^1\pm k_{1(2)}^2$. The quantity $M^\ast$ satisfies the condition 
\be 
M^{\ast2} = \frac{m_u^{\ast2} + \bfk^2}{x} + \frac{m_{\bar{d}}^{\ast2} + \bfk^2}{1-x} \, .
\ee 
The spin wave function given in Eq. (\ref{SpinWfns}) must satisfy the normalization condition
\be 
\sum_{\lambda_1 \lambda_2} \Phi^\ast (x,\bfk,\lambda_1,\lambda_2) \, \Phi(x,\bfk,\lambda_1,\lambda_2) = 1 \, .
\ee
We have adopted the Brodsky-Huang-Lepage prescription to define momentum space wave function
\cite{Kaur:2020vkq,Xiao:2002iv,Yu:2007hp} 
which is represented as follows
\be 
\varphi (x,\bfk) &=& \mathcal{A} \, exp \,\Biggl[-\frac{ \frac{m_{u}^{\ast2} + \bfk^2}{x} + \frac{m_{\bar{d}}^{\ast2} + \bfk^2}{1-x}}{8 \beta^2} \nonumber \\ &-& \frac{(m_u^{\ast2} - m_{\bar{d}}^{\ast2})^2}{8 \beta^2 \, \bigg( \frac{m_u^{\ast2} + \bfk^2}{x} + \frac{m_{\bar{d}}^{\ast2} + \bfk^2}{1-x}\bigg)}\Biggr] \, ,
\ee 
where $\mathcal{A}= A \, exp \, \big[\frac{m_u^{\ast2} + m_{\bar{d}}^{\ast2}}{8 \beta^2}\big]$ with $A$ and $\beta$ representing the normalization constant and harmonic scale parameter respectively. The momentum space wave function is normalized as
\begin{eqnarray}
    \int \frac{{d x} d^2 \bfk}{2 (2 \pi)^3} \, |\varphi (x,\bfk)|^2 =1 \, .
\end{eqnarray}

\par The in-medium longitudinal momentum fraction ($x^*$) of $j$th quark/antiquark is related to vacuum longitudinal momentum fraction ($x$) by the relations \cite{Puhan:2024xdq}
 \begin{align}
 x_j^*  = 
 \begin{cases}
 \frac{E_j^* + g_{\omega}^{j}\omega + 
 g_{\rho}^{j} I^{3j}\rho + k_j^{*3}}{E_j^* + E_{\bar j}^* + g_{\rho}^{{ j}} \left(I^{3{j}}-I^{3{\bar j}}\right)\rho + P^{*3}} = \frac{x_j+ (g_{\omega}^{j}\omega + 
 g_{\rho}^{j} I^{3j}\rho)/P^+}{1+\left(I^{3{j}}-I^{3{\bar j}}\right)\rho/P^+} \quad \text{for quark } q \\
  \frac{E_{\bar j}^* - g_{\omega}^{j}\omega - 
  g_{\rho}^{j} I^{3\bar j}\rho + k_{\bar j}^{*3}}{E_j^* + E_{\bar j}^* + g_{\rho}^{{ j}} \left(I^{3{j}}-I^{3{\bar j}}\right)\rho + P^{*3}} = \frac{x_j - (g_{\omega}^{j}\omega + 
 g_{\rho}^{j} I^{3j}\rho)/P^+}{1+\left(I^{3{j}}-I^{3{\bar j}}\right)\rho/P^+} \quad
  \text{for antiquark } \bar{q} .
 \end{cases}
 \label{Eq_xfrac_med2}
 \end{align}
However, for the sake of simplicity, we have constrained our results to $x$ only.

\section{Transverse momentum parton dependent distributions (TMDs)}
\label{Sec_TMDs}
For spin-$0$ pseudo-scalar mesons, there are only two quark TMDs at the leading twist \cite{Meissner:2008ay}. These are unpolarized  $f_1(x,\bfk^2)$ TMDs (T-even) and Boer-Mulders $h^{\perp}_1(x,\bfk^2)$ TMD (T-odd). In this work, we have calculated the unpolarized $f_1(x,\bfk^2)$ quark TMD by solving the quark-quark distribution correlation function appearing in
SIDIS is defined as \cite{Puhan:2023ekt, Meissner:2008ay, Pasquini:2008ax}

\be
f_1(x,\bfk^2) &=& \frac{1}{2} \int \frac{dz^- d^2 z_\perp}{2 (2 \pi)^3} e^{i \bfk . z} 
\langle \pi(P^+,\bfP;S)| \nonumber \\
&& \bar{\vartheta}(0) \gamma^+ W(0,z) \vartheta (z)|\pi(P^+,\bfP;S) \rangle |_{z^+=0} \, ,
\label{tmd}
\ee
where $z=(z^+,z^-,z_\perp)$ is the position four vector. $\vartheta (z)$ is the quark field operator and $W(0,z)$ is the Wilson line, which connects the two quark field operators. However, in this work, Wilson line is taken as unity. By substituting pion state wave function in Eq. (\ref{tmd}), the overlap form of $f_1(x,\bfk^2)$ is found to be 

\be 
f_1(x,\bfk^2) &=& \frac{1}{16 \pi^3} \\
&\times& \big[|\psi (x,\bfk,\uparrow,\uparrow)|^2 +|\psi (x,\bfk,\uparrow,\downarrow)|^2 \\
&+& |\psi (x,\bfk,\downarrow,\uparrow)|^2 + |\psi (x,\bfk,\downarrow,\downarrow)|^2 \big] \, .
\label{overlap}
\ee
In explicit form, we have

\be
f_1(x,\bfk^2) &=&  \frac{1}{16 \pi^3} \\
&\times& \bigg[\big((x {M}^*+m^*_{u})((1-x){M}^*+m^*_{\bar d})-\textbf{k}^2_\perp\big)^2 \\
&+& \big({M}^*+ m^*_u+m^*_{\bar d}\big)^2\bigg] \frac{\mid \varphi^*(x,\textbf{k}_\perp)\mid^2}{c^{*2}_1 c^{*2}_2}. 
\label{pdf}
\ee
Similarly, one can obtain the anti-quark TMD by implicating the relation 
\begin{eqnarray}
    f^q_1(x,\bfk^2) = f^{\bar q}_1(1-x,-\bfk^2),
    \label{rule}
\end{eqnarray}
to verify the conservation of momentum. The unpolarized PDF for pion at the leading twist can be calculated by integrating the $f_1(x,\bfk^2)$ TMD over transverse momentum of quark as \cite{Kaur:2020vkq, Maji:2016yqo}
\begin{eqnarray}
    f_1(x) =\int d^2 \bfk f_1(x,\bfk^2),
    \label{pdf}
    \label{rule}
\end{eqnarray}
This unpolarized $ f_1(x)$ PDF obeys all the PDF sum rule discussed in Refs. \cite{Puhan:2023ekt,Kaur:2020vkq}

\begin{figure*}
\centering
\begin{minipage}[c]{0.98\textwidth}
(a)\includegraphics[width=5.45cm]{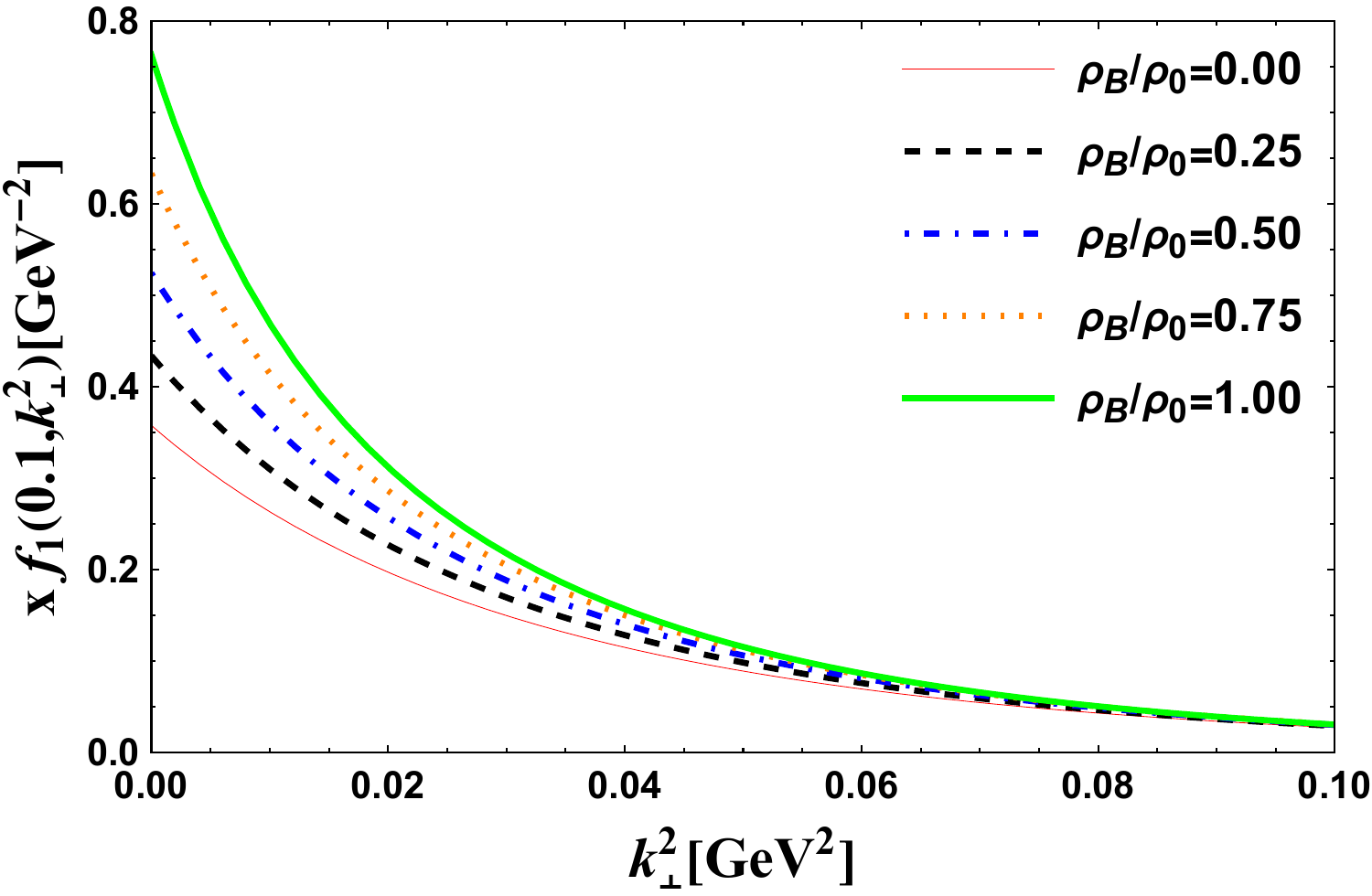}
\hspace{0.03cm}	
(b)\includegraphics[width=5.45cm]{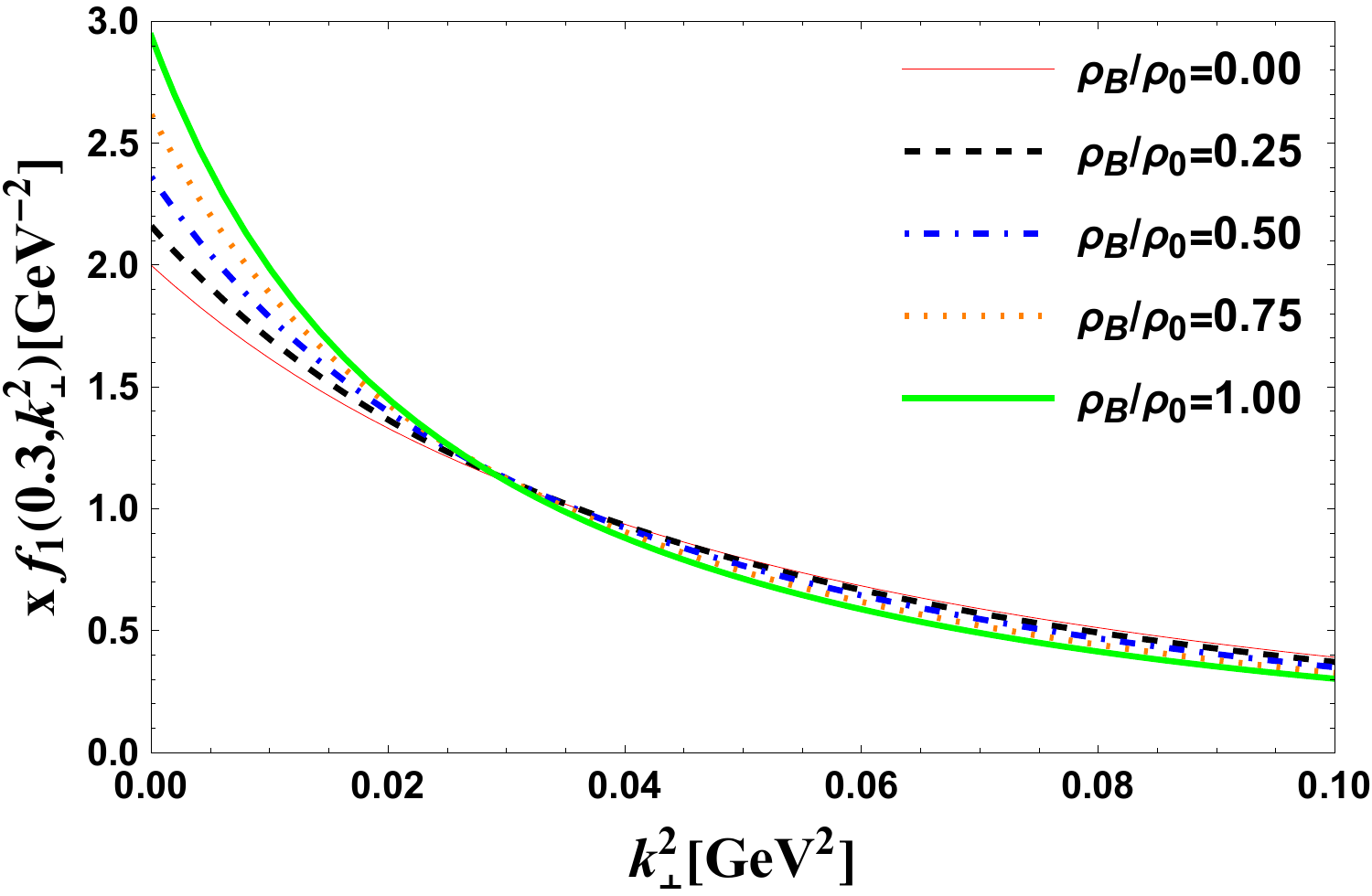}
\hspace{0.03cm}
(c)\includegraphics[width=5.45cm]{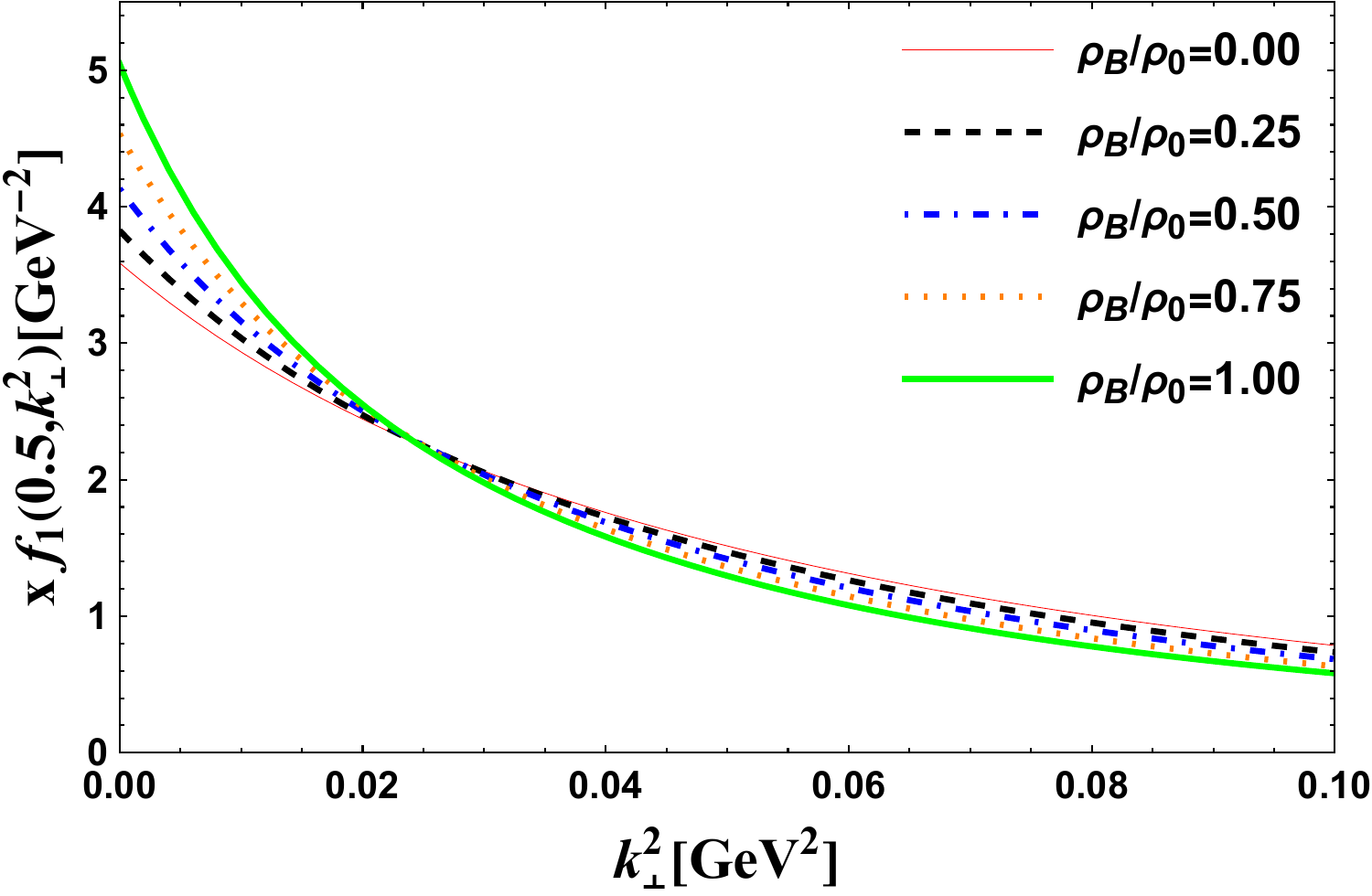}
\hspace{0.03cm} \\
(d)\includegraphics[width=5.45cm]{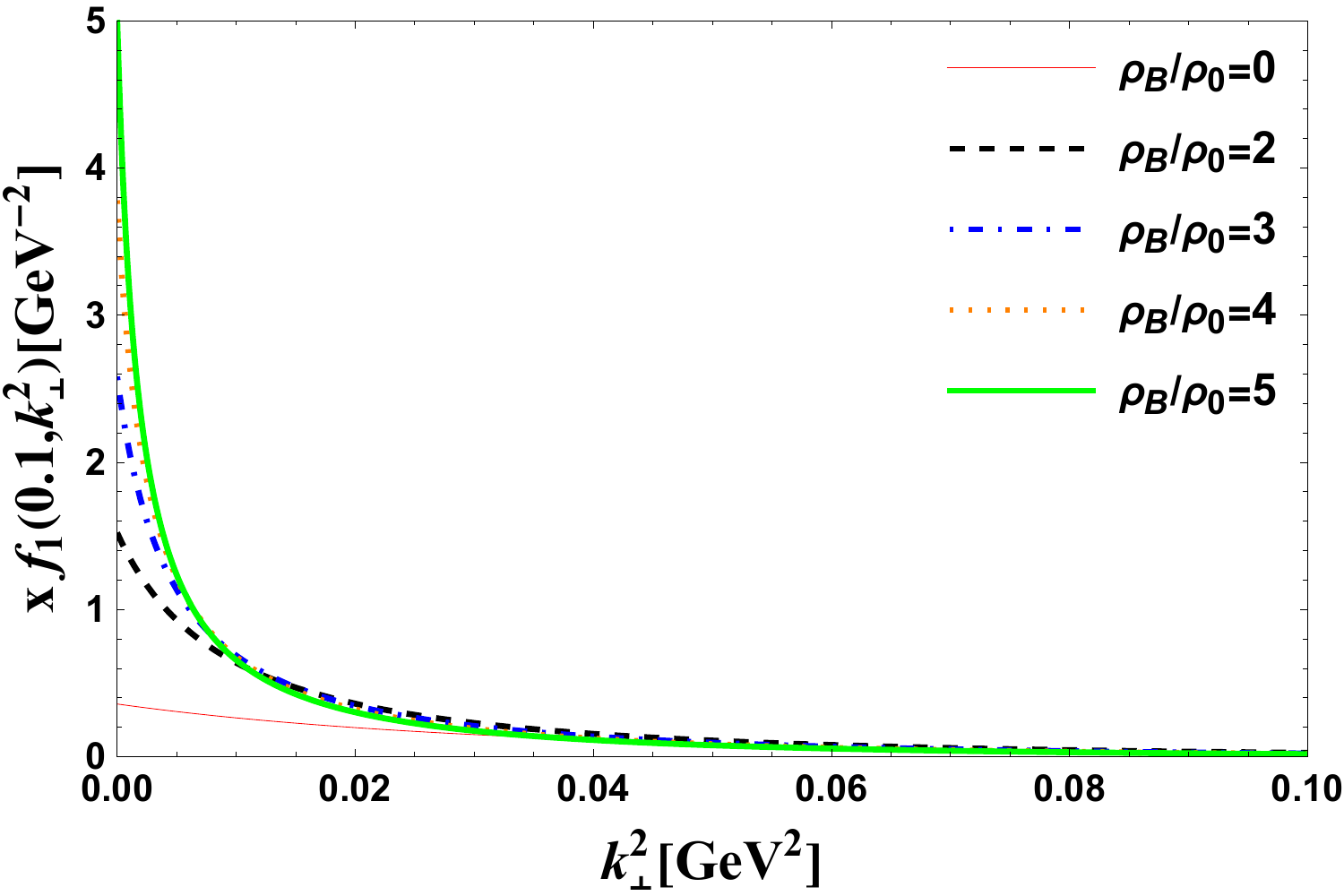}
\hspace{0.03cm}	
(e)\includegraphics[width=5.45cm]{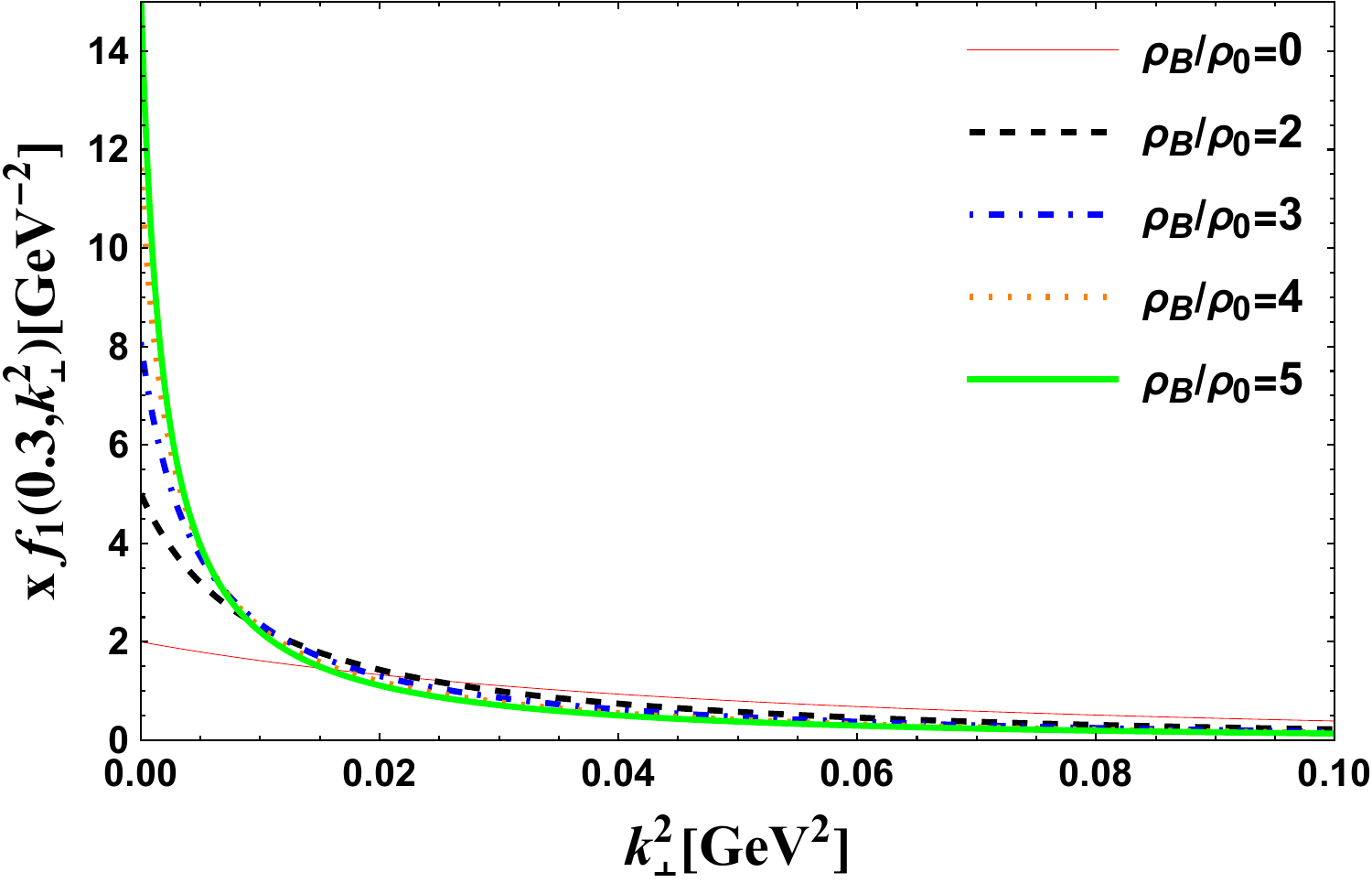}
\hspace{0.03cm}
(f)\includegraphics[width=5.45cm]{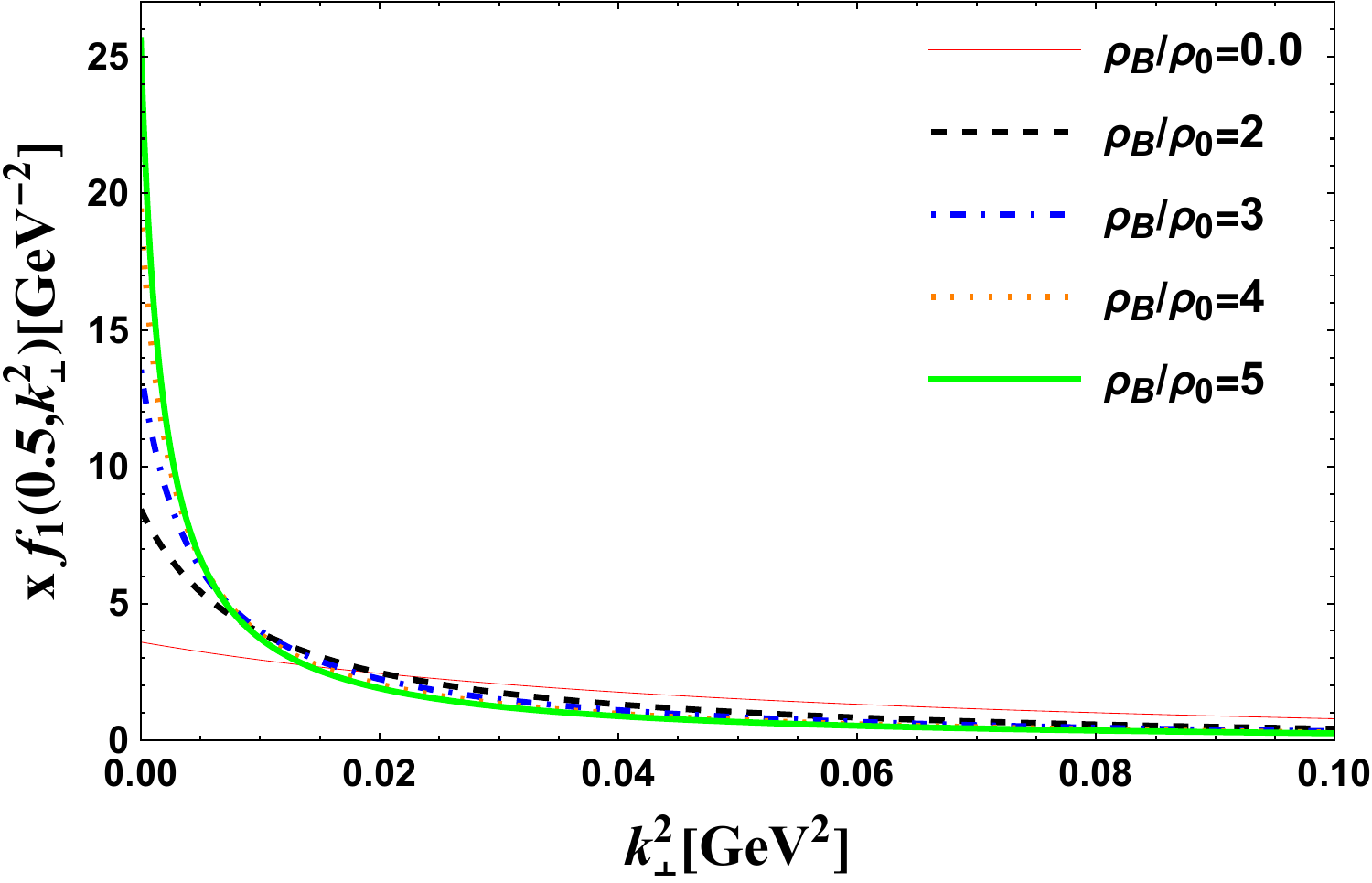}
\hspace{0.03cm}
\end{minipage}
\caption{\label{TMDsk} (Color online) Unpolarized T-even TMD, $x f(x,\bfk^2)$ as a function of quark's transverse momentum $\bfk$ for three different values of longitudinal momentum fraction $x=0.1$ (left panel), $0.3$ (central panel) and $0.5$ (right panel). First row represents the comparison of vacuum and in-medium unpolarized T-even TMDs upto  baryonic density $\rho_B=\rho_0$ with a step size of $0.25$, whereas second row represents the comparison of vacuum and in-medium unpolarized T-even TMDs with baryonic density above $\rho_B=\rho_0$ upto $\rho_B=5 \rho_0$ with a step size of $1$.}
\end{figure*}
\begin{figure*}
\centering
\begin{minipage}[c]{0.98\textwidth}
(a)\includegraphics[width=5.45cm]{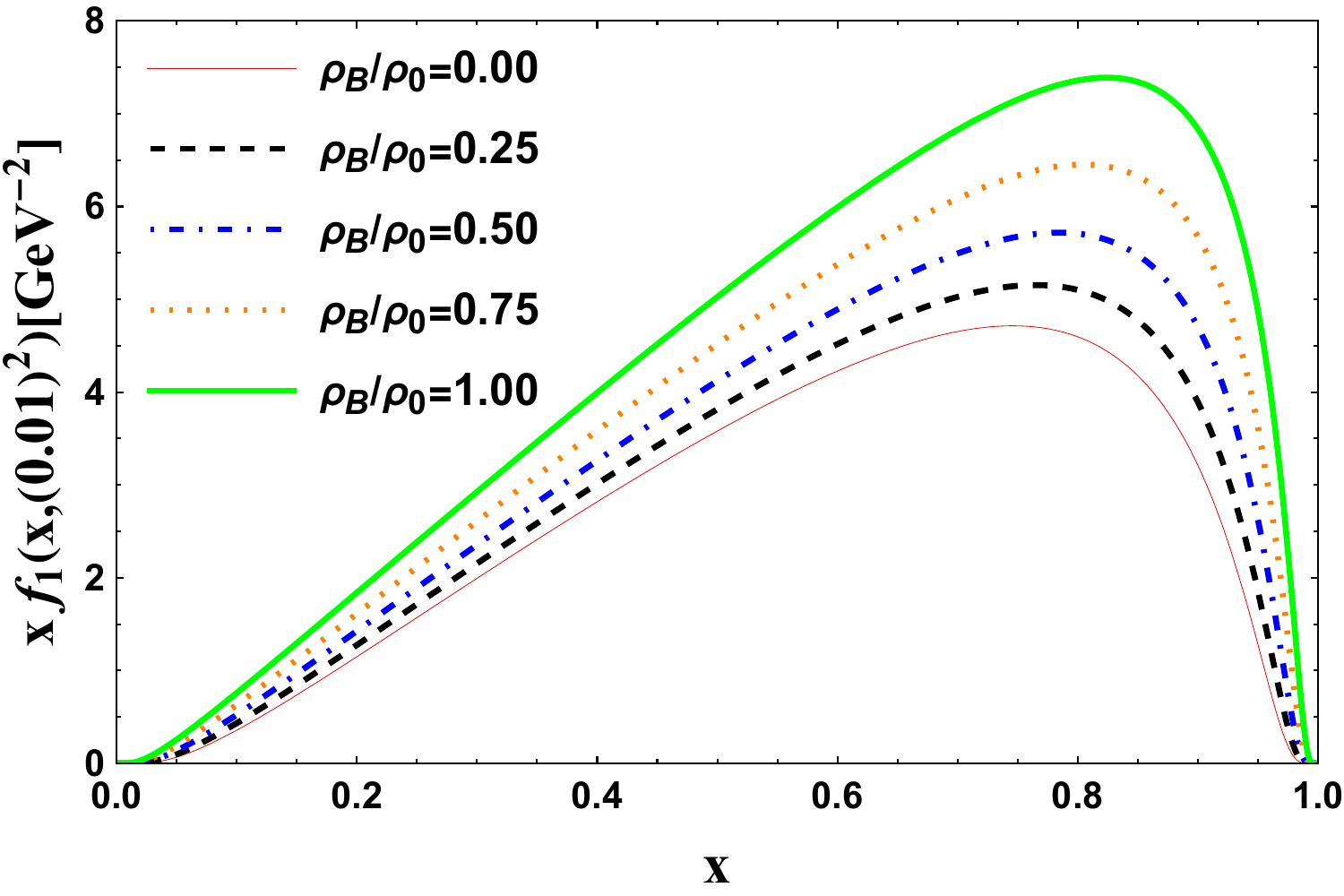}
\hspace{0.03cm}	
(b)\includegraphics[width=5.45cm]{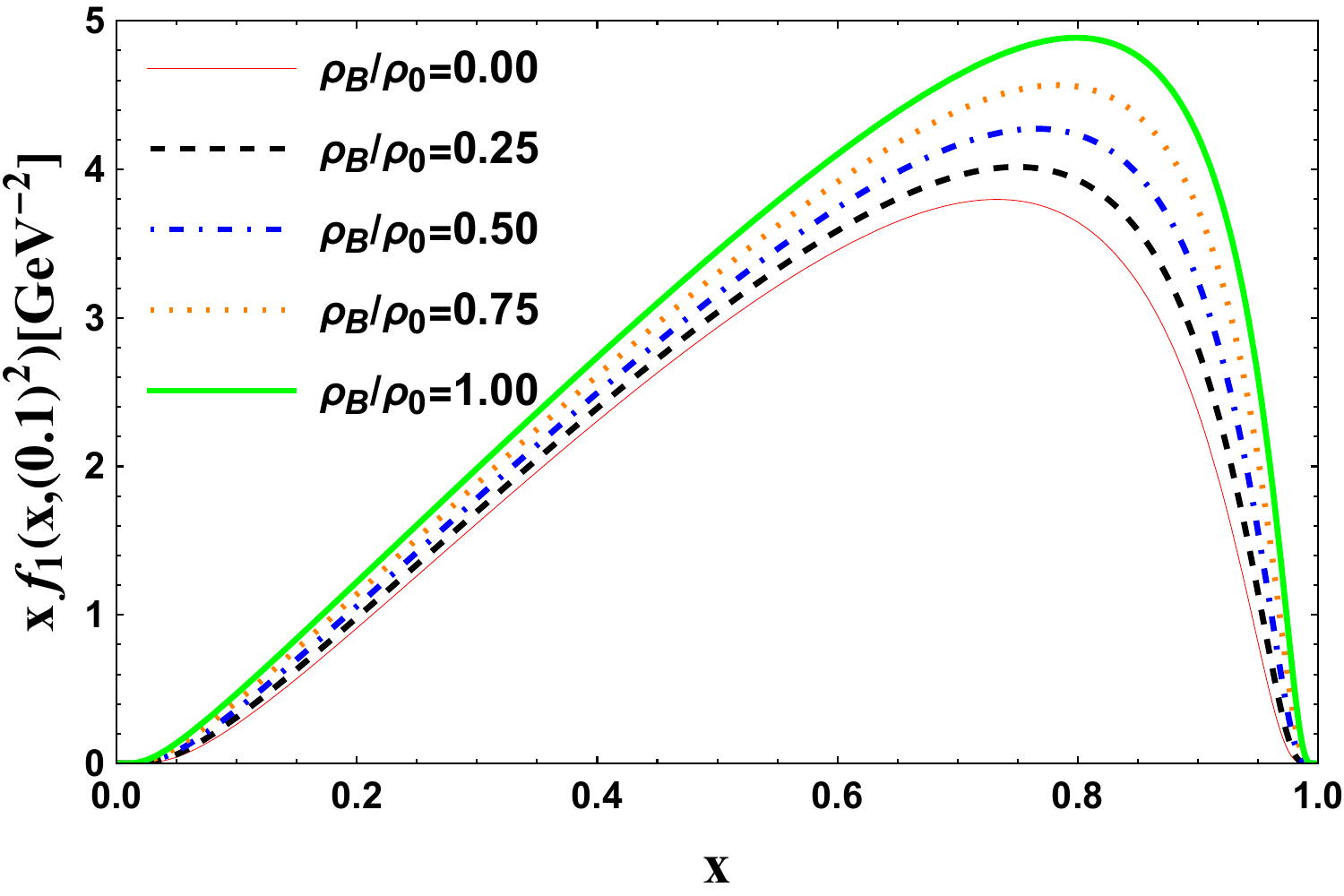}
\hspace{0.03cm}
(c)\includegraphics[width=5.45cm]{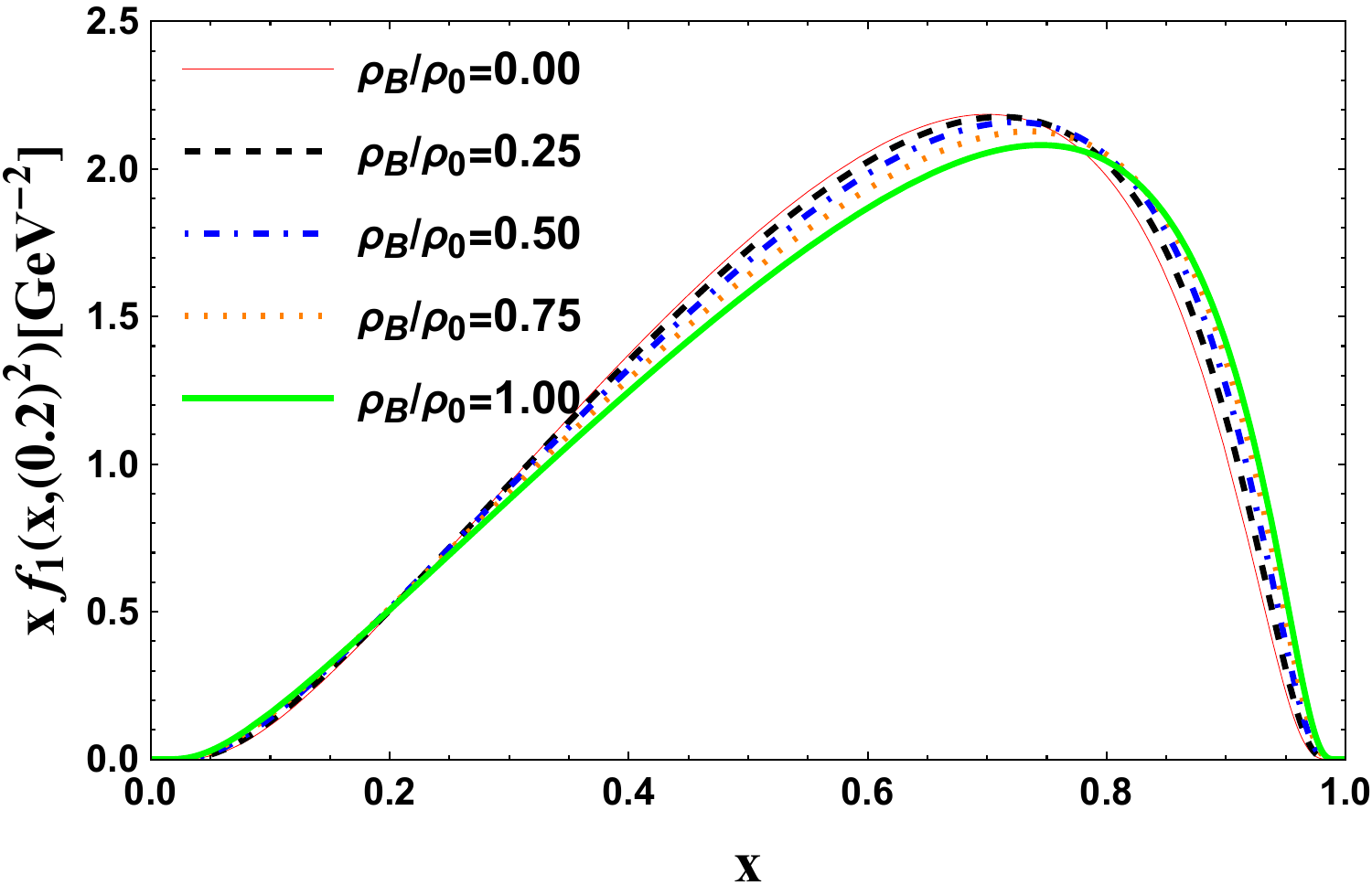}
\hspace{0.03cm} \\
(d)\includegraphics[width=5.45cm]{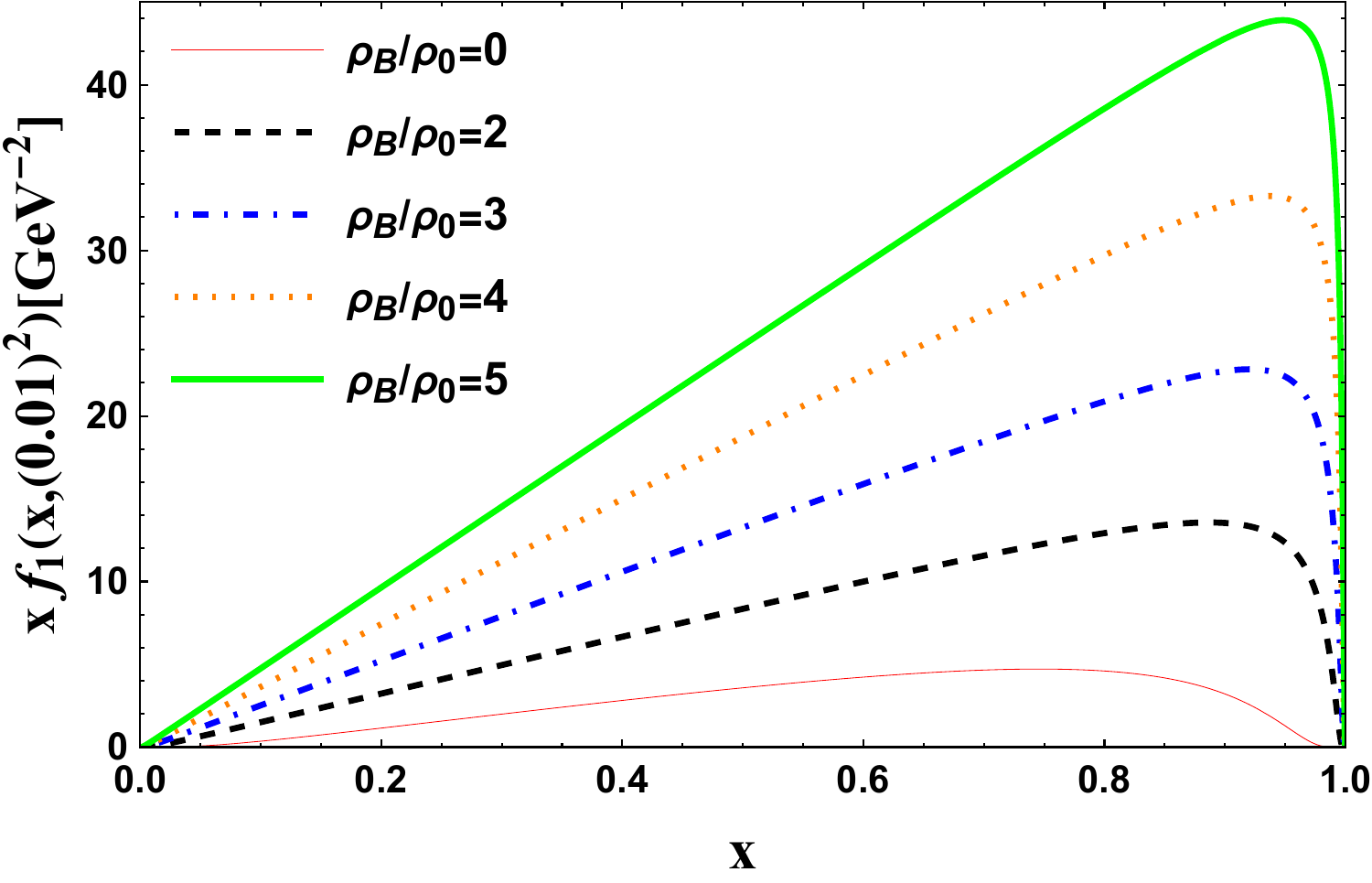}
\hspace{0.03cm}	
(e)\includegraphics[width=5.45cm]{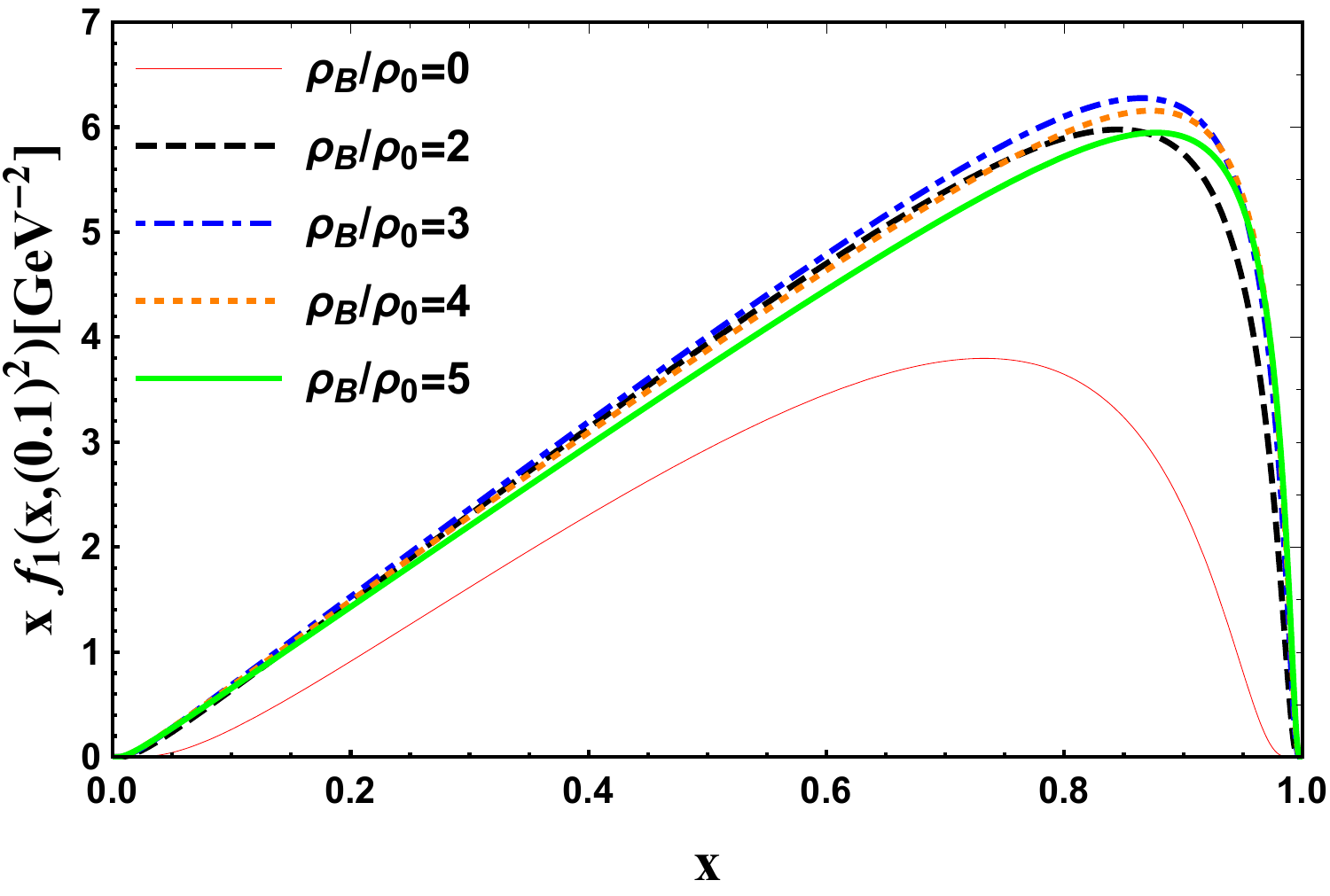}
\hspace{0.03cm}
(f)\includegraphics[width=5.45cm]{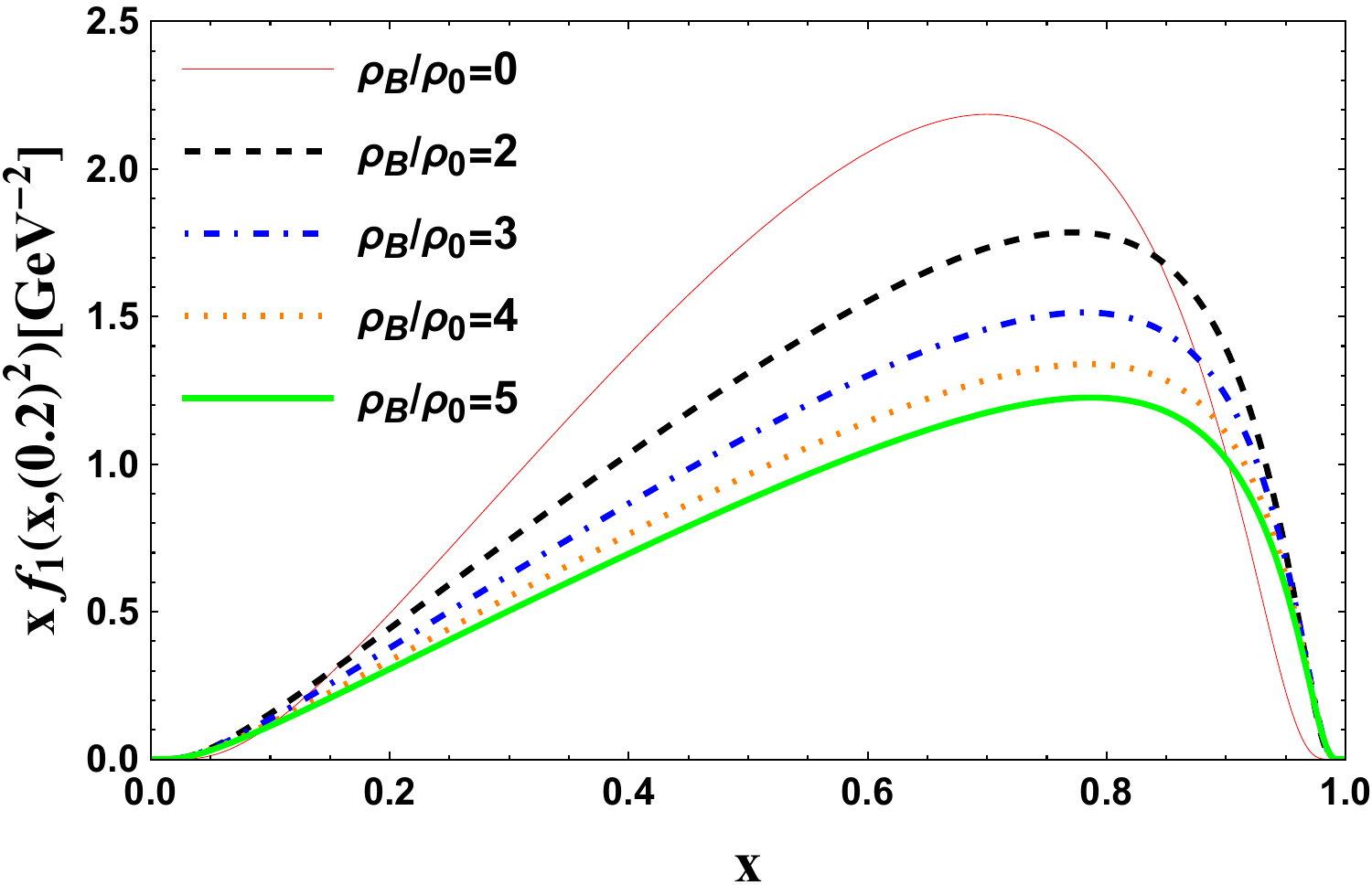}
\hspace{0.03cm}
\end{minipage}
\caption{\label{TMDsx} (Color online) Unpolarized T-even TMD, $x f(x,\bfk^2)$ as a function of longitudinal momentum fraction $x$ for three different values of quark's transverse momentum $\bfk=0.01$ GeV (left panel), $0.1$ GeV (central panel) and $0.2$ GeV (right panel). First row represents the comparison of vacuum and in-medium unpolarized T-even TMDs upto  baryonic density $\rho_B=\rho_0$ with a step size of $0.25$, whereas second row represents the comparison of vacuum and in-medium unpolarized T-even TMDs with baryonic density above $\rho_B=\rho_0$ upto $\rho_B=5 \rho_0$ with a step size of $1$.}
\end{figure*}
\section{Results and discussion}
\label{Sec_results}
For numerical calculations, we need only two input parameters: quark masses ($m_{u(d)}$) and harmonic scale constant ($\beta$). In this work, the quark masses get modified in the  isospin asymmetric nuclear matter for different nuclear densities. These parameters have been fixed using pion masses and decay constant.
In Fig. \ref{TMDsk}, the distributions of TMDs with respect to transverse momentum $\bfk^2$ carried by an active quark have been presented for fixed values of longitudinal momentum fraction $x$ on wide range of baryonic density. Fig. \ref{TMDsk} (a) illustrates the comparative distribution of vacuum and in-medium TMDs for baryonic density upto $\rho_B=\rho_0$ with a step size of $0.25$ at $x=0.1$ and portrays similar trend of smooth fall-off with increase in $\bfk^2$. At lower values of $\bfk^2$, higher the baryonic density, higher is the amplitude of the distribution. However, in case of $x=0.3$ and $x=0.5$, the trend of enhancement in the amplitude of the distributions with increase in baryonic density shows a reverse trend beyond $\bfk^2= 0.029$ GeV$^2$ and $\bfk^2=0.025$ GeV$^2$ respectively, after which the amplitude of the distributions decreases with an increase in baryonic density. This is clear form Fig. \ref{TMDsk} (b) and Fig. \ref{TMDsk} (c). The impact of different values of longitudinal momentum fraction $x$ on TMDs with different baryonic density in Fig. \ref{TMDsk} (a)-(c) implies that as $x$ increases, the amplitude of the distributions increases with significant difference among the distributions at higher values of $\bfk^2$. For baryonic density above $\rho_B=\rho_0$, comparative distributions of vacuum and in-medium TMDs, presented in second row of Fig. \ref{TMDsk}, shows a fast decrement for in-medium TMDs with increase of $\bfk$ as compared to the vacuum TMD as their amplitude of distribution increases with a great margin. Beyond $\bfk=0.06$ GeV, the distribution saturates, whereas for baryonic density $\rho_B \le \rho_0$, no such satuartion was observed. The overall impact of $x$ on TMDs for higher baryonic density is same as that of lower baryonic density. The variations observed in the in-medium distributions as compared to the  vacuum distribution are due to scaling down of effective quark masses with enhancement of baryonic density of matter which might be due to partial restoration of chiral symmetry.
The vacuum and in-medium results of unpolarized $x f_1(x)$ PDFs have been demonstrated in our previous work \cite{Puhan:2024xdq}, along with comparison of  our
vacuum ($\rho_B = 0$) and in-medium ($\rho_B = \rho_0$) evolved pion PDFs (from our model scale $Q_0^2=0.23$ GeV$^2$ to $Q^2=16$ GeV$^2$ by using the next-to-leading order (NLO) Dokshitzer-Gribov-Lipatov-Altarelli-Parisi (DGLAP) evolution equations) with E615 data \cite{Conway:1989fs} and modified
E615 data \cite{Aicher:2010cb} that showed a good agreement. The impact of asymmetry parameter $\eta$ was also found to be less on the distributions (as shown in Ref. \cite{Puhan:2024xdq}) as we are dealing only with the light quarks of pions. Hence, we present our distributions for the case of $\eta=0$ only. 

The impact of baryonic density on TMDs as a function of longitudinal momentum fraction $x$ carried by an active quark for fixed values of its transverse momentum is demonstrated in Fig. \ref{TMDsx}. For $\bfk=0.01$ GeV and baryonic density ranging from $0$ to $1$ with a difference of $0.25$ in the unit of $\rho_0$, the TMD distribution is presented in Fig. \ref{TMDsx} (a) which implies that as baryonic density increases, the amplitude of the distribution enhances with shifting of the peak towards higher value of $x$. Hence, higher dense is the matter, higher is the longitudinal momentum fraction carried by an active quark. As the transverse momentum increases to $\bfk=0.1$ GeV, a shift of peaks towards lower values of $x$ has been observed with decrease in the amplitude of distribution in Fig. \ref{TMDsx} (b). On further increasing transverse momentum to $\bfk=0.1$ GeV, an opposite behavior has been observed for the middle range of $x$ than other two distributions for $\bfk=0.01$ GeV and $\bfk=0.1$ GeV, presented in Fig. \ref{TMDsx} (c). Second row of Fig. \ref{TMDsx} represents the distributions corresponding to the baryonic density $2 \rho_0 \le \rho_B \le 5 \rho_0$ with vacuum TMD for the sake of  comparison. A significant shoot up in the amplitude of distributions has been observed in Fig. \ref{TMDsx} (d) for $\bfk=0.01$ GeV as compared to the in-medium distributions with baryonic density $\rho_B \le \rho_0$. However, for the case of $\bfk=0.1$ GeV, the amplitude of distributions increases upto $\rho_B=3 \rho_0$. A reverse impact on the amplitude of distributions has been observed for baryonic density $4 \rho_0$ and $5 \rho_0$. Further increase in the value of $\bfk$ decreases the amplitude of distribution towards higher values of $x$. In nut shell, higher the density of matter, lower will be the effective mass of an active quark and higher will be the longitudinal momentum fraction carried by it.  

\begin{figure*}
\centering
\includegraphics[width=5.45cm]{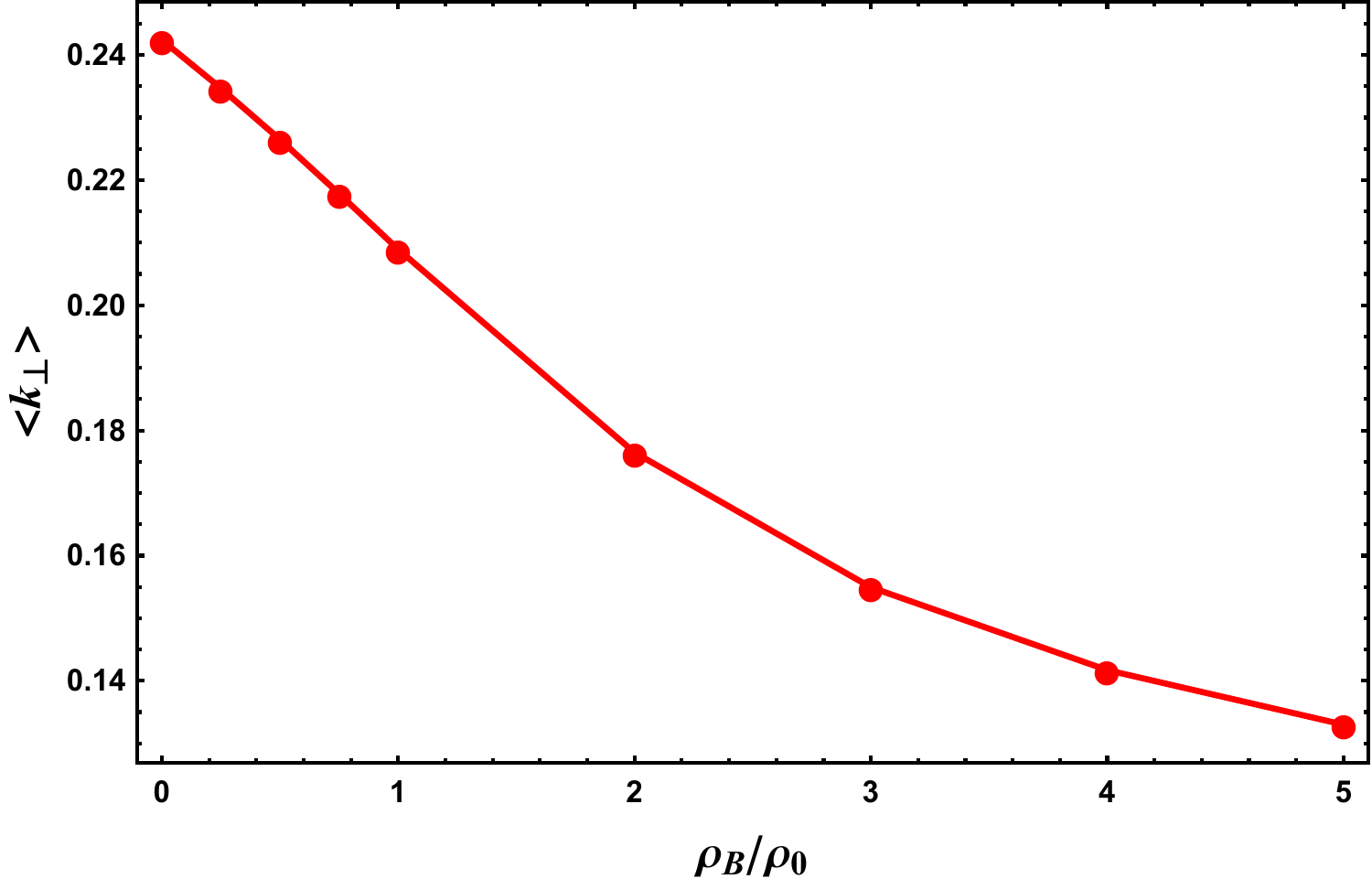}
\caption{\label{TMDaverage} (Color online) The average transverse momenta carried by the quark at different baryonic densities.}
\end{figure*}

\begin{figure*}
\centering
\begin{minipage}[c]{0.98\textwidth}
(a)\includegraphics[width=5.45cm]{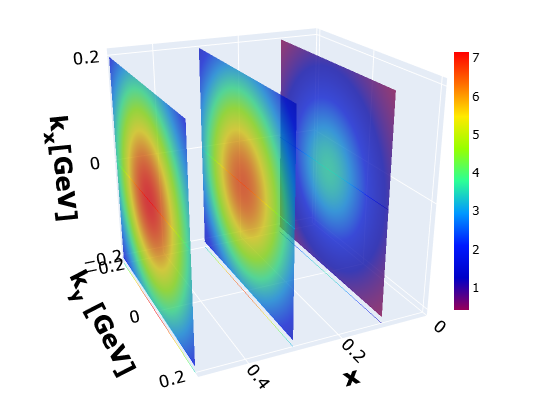}
\hspace{0.03cm}	
(b)\includegraphics[width=5.45cm]{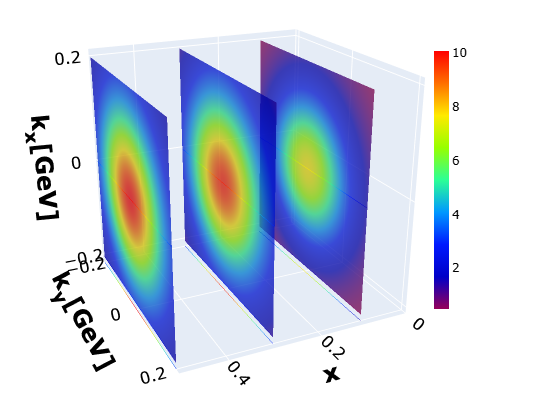}
\hspace{0.03cm}
(c)\includegraphics[width=5.45cm]{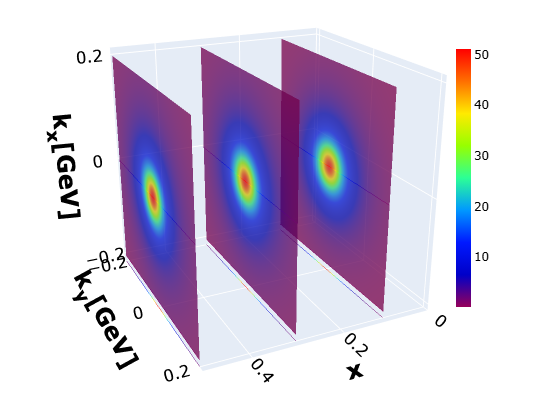}
\hspace{0.03cm}
\end{minipage}
\caption{\label{TMDskx} (Color online) Spin densities as a function of longitudinal momentum fraction $x$ in transverse plane of momentum.}
\end{figure*}

In Fig. \ref{TMDaverage}, the average momenta carried by an active quark of a pion is presented for different values of baryonic densities and it follows that beyond $\rho_B=2 \rho_0$, the rate of decrease in average momenta carried by an active quark slows down. Fig. \ref{TMDskx} represents the vacuum and in-medium transverse structure as spin densities of an active $u$ quark in $k_x k_y$ plane for three different values of longitudinal momentum fraction $x$. As TMD $f(x,\bfk^2)$ is unpolarized, so its transverse structure also comes out to be symmetric in nature. For the case of vacuum spin density, with increase in the value of $x$, spread of distribution increases with enhancement of its amplitude, as presented in Fig. \ref{TMDskx} (a). Whereas in case of pion immersed in matter with baryonic density $\rho_B=\rho_0$, the trend of spreading of transverse distribution with $x$ is same as for vacuum. However, the transverse spread is more confined towards centre with comparatively large amplitude as presented in \ref{TMDskx}  
(c). On further increasing the baryonic density to $\rho_B=5 \rho_0$, in Fig. \ref{TMDskx}, it is found that an active quark is more confined at the centre and follows similar trend as seen for the vacuum distributions with respect to $x$.



\section{Summary and conclusions}
\label{Sec_summary}
In this paper, we have investigated the in-medium valence quark transverse momentum-dependent distributions (TMDs) of the lightest pseudo-scalar meson, pion, immersed in an isospin asymmetric nuclear matter at zero temperature. The vacuum and in-medium TMDs have been studied using a light-cone quark model. The input parameters such as effective masses, required to induce medium effects in TMDs are obtained by employing chiral SU($3$) quark mean field model. Comparative analysis of vacuum and in-medium distributions are studied for baryonic density upto $\rho_B=5 \rho_0$.
The distributions of TMDs with respect to transverse momentum of an active quark, for fixed values of longitudinal momentum fraction $x$, have been analyzed. It is observed that higher the baryonic density, larger is the amplitude of the distribution over the entire $\bfk$ range. However, this is true only for $x=0.1$. Whereas, for $x=0.3$ and $x=0.5$, the trend is same upto the values of $\bfk^2= 0.029$ GeV$^2$ and $\bfk^2=0.025$ GeV$^2$ respectively. Beyond these values of $\bfk$, the trend gets reversed. The distributions of TMDs with respect to longitudinal momentum fraction $x$ of an active quark, for fixed values of transverse momentum implies that as the baryonic density increases, the amplitude of the distribution increases along with a  drift towards higher values of $x$. This drift is consistent for all the values of baryonic densities and transverse momenta. However, in this case also a reverse trend is observed in the behavior of amplitude after a certain value of baryonic density and transverse momentum. At higher density of baryonic matter, the average transverse momenta are found to decrease rapidly. The momentum space representation of spin densities for an active $u$ quark in the transverse plane of a moving pion shrinks towards the center of pion, when immersed in the isospin asymmetric nuclear matter. These changes may be attributed to scaling down of quark masses which infer the partial restoration of chiral symmetry.

\section*{Acknowledgements}

H.D. would like to thank  the Science and Engineering Research Board, Anusandhan-National Research Foundation, Government of India under the scheme SERB-POWER Fellowship (Ref No. SPF/2023/000116) for financial support.

\appendix



\end{document}